\documentstyle[12pt]{article}
\author{V. A. Tsokur, Yu. M. Zinoviev
        \thanks{E-mail address: ZINOVIEV@MX.IHEP.SU}\\
        {\it Institute for High Energy Physics } \\
        {\it Protvino, Moscow Region, 142284, Russia}}
\title{Spontaneous supersymmetry breaking \\
       in $N=3$ supergravity with matter}
\date{}
\begin{document}
\maketitle

\thispagestyle{empty}

\begin{abstract}
In this paper we investigate the problem of spontaneous supersymmetry
breaking without a cosmological term in $N=3$ supergravity with matter
vector multiplets, scalar fields geometry being
$SU(3,m)/SU(3)\otimes SU(m)\otimes U(1)$. At first, we consider the case
of minimal
coupling with different possible gaugings (compact as well as
non-compact). Then we show that there exist dual version of such a
theory (with the same scalar field geometry), which turns out to be
the generalization of the $N=3$ hidden sector, constructed some time
ago by one of us, to the case of arbitrary number of vector
multiplets. We demonstrate that spontaneous supersymmetry breaking is
still possible in the presence of matter multiplets.
\end{abstract}

\newpage
\setcounter{page}{1}

\section{Introduction}

   The problem of spontaneous supersymmetry breaking without a
cosmological term still remains to be one of the most important and
hard ones in supergravity as well as in superstring theories. In this
paper we are not going to discuss this problem in the superstrings, but
we feel that the lack of complete understanding of the spontaneous
supersymmetry breaking in extended supergravities may be (at least one
of) the reason, why the appropriate solution of this problem in
superstrings is still missing. Only in $N=1$ supergravity, due to
large (functional) arbitrariness in general coupling of vector and
chiral multiplets, this problem has been successfully solved, which
make possible a lot of phenomenological applications.

   In extended supergravities this task appeared to be much more
complicated. One of the reasons is that a very high symmetry such
theories possess leads to a severe restriction on the possible matter
couplings (but at the same time could lead to the models with very
high predictive power). The second one is the necessity to have a
spontaneous supersymmetry breaking with at least two different scales,
when extended supersymmetry is broken to the $N=1$ at the first stage.
For example, for $N=2$ supergravity a wide class of so called no-scale
models has been constructed \cite{Wit85,Cre85,Ito87}, in which two
supersymmetries were broken with only one scale and the gravitini
were mass degenerate. At the same time, the realization of partial
super-Higgs effect $N=2 \to  N=1$ appeared to be non-trivial task
\cite{Cec86}. Later on, the $N=2$ hidden sector, consisting of $N=2$
supergravity itself, one vector and one hypermultiplet and admitting
spontaneous supersymmetry breaking with two arbitrary scales, has
been constructed \cite{Zin86}. At first sight, it seemed that the only
natural generalization of such hidden sector to the case of arbitrary
number of vector multiplets was the model with the scalar field
geometry $SO(2,m)/SO(2)\otimes SO(m)$. The spontaneous supersymmetry
breaking
without a cosmological term was indeed shown to be possible in such
theories and the examples of corresponding soft breaking terms have
been calculated \cite{Zin86a,Zin90}. At last, recently it was shown
\cite{Zin94}, that there exist a dual version of the $N=2$
supergravity --- vector multiplets system with the scalar field
geometry $SU(1,m)/SU(m)\otimes U(1)$ which corresponds to different
interaction of arbitrary number of vector multiplets with the same
hidden sector, thus leading to the same possibilities for spontaneous
supersymmetry breaking.

  In this paper we (re)consider the problem of spontaneous
supersymmetry breaking in $N=3$ supergravity with vector multiplets,
while our subsequent paper deals with the $N=4$ case. Firstly, we
recall the construction ot the "minimal" coupling, scalar field
geometry being $SU(3,m)/SU(3)\otimes SU(m)\otimes U(1)$ and describes
different
gaugings possible in such model. Then, using the fact that the dual
versions have the same scalar field geometry, we have managed to
construct a model, which corresponds to the interaction of arbitrary
number of vector multiplets with $N=3$ hidden sector, admitting
spontaneous supersymmetry without cosmological term and with three
different scales. We have shown that spontaneous supersymmetry
breaking remains to be possible in the presence of matter multiplets.

\section{$N=3$ supergravity with vector multiplets}

  In this section we investigate an "ordinary" version of $N=3$
supergravity interacting with vector multiplets. As is well known,
scalar fields in extended supergravities describe, as a rule,
non-linear $\sigma $-models of the type $G/H$, where $G$ is some
non-compact
group, and $H$ --- its maximal compact subgroup.  As scalar fields of
$N=3$ vector multiplets are transformed under the triplet
representation of $SU(3)$ group, a $\sigma $-model
$SU(3,m)/SU(3)\otimes SU(m)\otimes U(1)$ seems to be the (only) natural
candidate
for the scalar field geometry in this case. Indeed, such a model has
been constructed some time ago by one of us \cite{Zin86b} (see also
\cite{Cas86}). For our discussion here to be self-contained, we
reproduce all the necessary formulas below.

  Let us consider the following fields: graviton $e_{\mu r}$, gravitini
$\Psi _{\mu i}$, $i=1,2,3$, majorana spinor $\rho $ and $m+3$ vector
multiplets
with vector fields $A_\mu {}^A$, majorana spinors $\Omega _{iA}$ and
$\lambda ^A$ and
complex scalar fields $z_i{}^A$. Here $A = 1,2...m+3$ with $g_{AB} =
diag(---,+...+)$. The supertransformations for the free vector
multiplets have the form:
\begin{eqnarray}
 \delta A_\mu {}^A &=& i(\bar{\Omega }^{iA} \gamma _\mu  \eta _i) \qquad
\delta \lambda ^A= - i
\hat{\partial } \bar{z}^i{}_{A} \eta _i \nonumber \\
 \delta \Omega _{iA} &=& - \frac{1}{2} (\sigma A)^A \eta _i + i
\varepsilon _{ijk} \hat{\partial } z_j{}^A \eta _k
\nonumber \\
 \delta \varphi _i{}^A &=& (\bar{\lambda }^A \eta _i) + \varepsilon _{ijk}
(\bar{\Omega }^{jA} \eta ^k) \label{n31}
\\
 \delta \pi _i{}^A &=& - (\bar{\lambda }^A \gamma _{5} \eta _i) +
\varepsilon _{ijk} (\bar{\Omega }^{jA} \gamma _{5}
\eta ^k) \nonumber
\end{eqnarray}
As it can easily be noted, the set of spinor and scalar fields is
superfluous, that allows one to have a symmetric description of three
vector fields of $N=3$ supergravity multiplet and $m$ vector fields
from matter multiplets. Therefore, we impose the following invariant
constraints on the scalar fields, corresponding to the geometry
$SU(3,m)/SU(3)\otimes SU(m)\otimes U(1)$:
\begin{equation}
 z_i{}^A \bar{z}^j{}_A = - 2 \delta _i{}^j \label{c1}
\end{equation}
(in the system of units, where gravitational constant $k=1$). For these
constraints to be consistent with supersymmetry we must impose
appropriate ones on the spinor fields, namely:
\begin{equation}
 z_i \Omega _j = 0 \qquad \bar{z}^i \lambda  = 0 \label{c2}
\end{equation}
Here and further on, we shall omit, wherever possible, the repeated
indices. Besides, we use $\gamma $-matrix representation in which majorana
spinors are real. Therefore in all expressions with spinors
$\gamma _5$-matrix will play a role of imaginary unit, for example, $z_i
\Omega _j
= (x_i + \gamma _5 y_i) \Omega _j$ and so on.

  In turn, the requirement of consistency for this constraints leads
to two important consequences. Firstly, the whole theory must be
invariant under the local $SU(3)\otimes U(1)$ transformations, in this, a
combination $(z_i \partial _\mu  \bar{z}^i)$ plays the role of a gauge
field. Let
us give here necessary covariant derivatives:
\begin{eqnarray}
 D_\mu  z_i{}^A &=& \partial _\mu  z_i{}^A - \frac{1}{2} (z_i \partial
_\mu  \bar{z}^j)
z_j{}^A \qquad \bar{z}^i D_\mu  z_j = O \nonumber \\
 D_\mu  \Omega _{iA} &=& D^g_\mu  \Omega _{iA} - \frac{1}{2} (z_i \partial
_\mu  \bar{z}^j) \Omega _{jA} +
\frac{1}{4} (z \partial _\mu  \bar{z}) \Omega _{iA}   \\
 D_\mu  \lambda ^A &=& D^g_\mu  \lambda ^A - \frac{1}{4} (z \partial _{\mu
} \bar{z}) \lambda ^A
 \nonumber
\end{eqnarray}
Note, that axial $U(1)_A$ charges of all fields are unambiguously
determined by the form of supertransformations.

  Secondly, in order to modify the spinors supertransformations to
make them consistent with the constraints, we shall need one more
important object. Let us introduce symmetric matrix $g_{ij} = z_i{}^A
z_j{}^A$. By virtue of constraints this matrix is nondegenerate (at
least in weak field approximation) therefore one can define an inverse
matrix $(g^{-1})^{ij}$, such that
\begin{eqnarray}
 (g^{-1})^{ij} z_j z_k &=& \delta ^i{}_k
\end{eqnarray}
It is not necessary to calculate matrix $(g^{-1})^{ij}$ explicitly
since it appears to be sufficient to use relations of the type:
\begin{eqnarray}
 \delta (g^{-1})^{ij} &=& - (g^{-1})^{ik} \delta (z_k z_l)(g^{-1})^{lj}
\end{eqnarray}

  In these notations the Lagrangian of interaction has the form:
\begin{eqnarray}
 L_o &=& - \frac{1}{2} R + \frac{i}{2} \varepsilon ^{\mu \nu \rho \sigma }
\bar{\Psi }_\mu {}^i
\gamma _5 \gamma _\nu  D_\rho  \Psi _{\sigma i} + \frac{i}{2} \bar{\rho }
\hat{D} \rho  + \frac{i}{2}
\bar{\Omega }^i \hat{D} \Omega _i + \frac{i}{2} \bar{\lambda } \hat{D}
\lambda  - \nonumber \\
 && - \frac{1}{4} A_{\mu \nu }{}^2 + \frac{1}{4} (g^{-1})^{ij} z_i{}^A
z_j{}^B A_{\mu \nu ,A} (A_{\mu \nu } + i \tilde{A}_{\mu \nu })_B+ h.c. +
\nonumber \\
 && + \frac{1}{2} D_\mu  z_i D^\mu  \bar{z}^i + \frac{i}{2\sqrt{2}}
\bar{\rho } \gamma ^\mu  (g^{-1})^{ij} z_i{}^A (\sigma A)_A \Psi _{\mu j}
+ \nonumber \\
 &&  + \frac{1}{2} \varepsilon ^{ijk} \bar{\Psi }_{\mu i}
(\bar{g}^{-1})_{jl}
\bar{z}^l{}_A (A^{\mu \nu } - \gamma _5 \tilde{A}^{\mu \nu })^A \Psi _{\nu
k} + \nonumber \\
 && + \frac{i}{4} \bar{\Omega }^{iA} \gamma ^\mu  \left[ (\sigma A)_A -
(g^{-1})^{jk}
z_j{}^A z_k{}^B (\sigma A)_B \right] \Psi _{\mu i} + \nonumber  \\
 && + \frac{1}{2} \varepsilon ^{ijk} \bar{\Omega }_{iA} \gamma ^\mu
\gamma ^\nu  D_\nu  z_j{}^A \Psi _{\mu k} -
\frac{1}{2} \bar{\lambda }^A \gamma ^\mu  \gamma ^\nu  D_\nu
\bar{z}^i{}_A \Psi _{\mu i} - \nonumber \\
 && - \frac{1}{2\sqrt{2}} \bar{\lambda }^A \left[ (\sigma A)^A -
(g^{-1})^{ij}
z_i{}^A z_j{}^B (\sigma A)^B \right] \rho  \nonumber \\
 &&  -  \frac{1}{2} \bar{\lambda }^A (g^{-1})^{ij} z_i{}^B (\sigma A)_B
\Omega _{jA}
 \label{n34}
\end{eqnarray}
Here and further on, we will omit the four-fermionic terms in our
Lagrangians (as well as bilinear in fermionic fields terms in the
supertransformations) which are unessential for the problem of
spontaneous supersymmetry breaking. This Lagrangian is invariant under
the following local supertransformations:
\begin{eqnarray}
 \delta e_{\mu r} &=& i(\bar{\Psi }_\mu {}^i \gamma _r \eta _i), \qquad
\delta \Psi _{\mu i} = 2D_\mu  \eta _i +
\frac{i}{2} \varepsilon _{ijk} (g^{-1})^{jl} z_l{}^A (\sigma A)_A \gamma
_\mu  \eta ^k \nonumber \\
 \delta A_\mu {}^A &=& - \varepsilon ^{ijk} (\bar{\Psi }_{\mu i} z_j{}^A
\eta _k) + \frac{i}
{\sqrt{2}} (\bar{\rho } \gamma _\mu  \bar{z}^i{}_A \eta _i) +
i(\bar{\Omega }^{iA} \gamma _\mu  \eta _i)
\nonumber \\
 \delta \rho  &=& -  \frac{1}{\sqrt{2}} (g^{-1})^{ij} z_i{}^A (\sigma A)^A
\eta _j \qquad
\delta \lambda ^A = - i \hat{D} \bar{z}^i{}_A \eta _i \label{n35}  \\
 \delta \Omega _{iA} &=& - \frac{1}{2} \left[ (\sigma A)^A - (g^{-1})^{jk}
z_j{}^A
z_k{}^B (\sigma A)^B \right] \eta _i + i \varepsilon _{ijk} \hat{D}
z_j{}^A \eta _k
\nonumber \\
 \delta \varphi _i{}^A &=& (\bar{\lambda }^A \eta _i) + \varepsilon _{ijk}
(\bar{\Omega }^{jA} \eta ^k), \qquad
\delta \pi _i{}^A = - (\bar{\lambda }^A \gamma _5 \eta _i) + \varepsilon
_{ijk} (\bar{\Omega }^{jA} \gamma _5 \eta ^k)
\nonumber
\end{eqnarray}

   Note, that we are working with physical fields only (without
introducing auxiliary ones), therefore for the fermionic fields the
algebra of the supertransformations closes on the mass shell only. At
the same time, for the bosonic fields this algebra closes without
using equations of motion, giving a nontrivial check of selfconsistency
for the model. For example, for the vector fields $A_\mu {}^A$ we have:
\begin{eqnarray}
 [\delta (\xi ), \delta (\eta )] A_\mu {}^A &=& - 2i (\bar{\eta }^i \gamma
^\nu  \xi _i) A_{\nu \mu }{}^A +
2 \partial _\mu  (\varepsilon ^{ijk} \bar{\eta }_i z_j{}^A \xi _k)
\label{n36}
\end{eqnarray}

    The Lagrangian (\ref{n34}) describes an interaction of $N=3$
supergravity with massless abelian vector multiplets. In extended
supergravities the appearance of the mass terms (and in general of the
scalar fields potential and the Yukawa type interactions) is
unambiguously connected with the switching of gauge interaction.
Besides supertransformations (\ref{n35}), the Lagrangian (\ref{n34})
is invariant under global transformations of the group $O(3,m)$
(maximal subgroup of $SU(3,m)$, not containing dual transformations of
vector fields). This allows one to introduce non-abelian gauge
interaction with the group $G \subset  O(3,m)$. For that purpose let us
replace the derivatives in formulas (\ref{n34}), (\ref{n35}) by
$G$-covariant ones, for example, $\partial _\mu  z_i{}^A \to  \partial
_\mu  z_i{}^A + f^{ABC}
A_\mu {}^B z_i{}^C$, where $f^{ABC}$ is structure constants of group
$G$. In order to compensate the violation of the invariance under the
supertransformations which arises after that, it is necessary to add
to the Lagrangian additional terms of the form:
\begin{eqnarray}
 L' &=& f^{ABC}\{ \frac{1}{8} \bar{\Psi }_{\mu i} \sigma ^{\mu \nu }
\varepsilon ^{ijk} z_j{}^A
z_k{}^B \bar{z}^l{}_C \Psi _{\nu l} - \frac{i}{2} \bar{\Psi }_\mu {}^i
\gamma ^\mu  z_i{}^A
\bar{z}^j{}_B \Omega _{jC} + \nonumber \\
 && + \frac{i}{4} \bar{\Psi }_\mu {}^i \gamma ^\mu  z^A \bar{z}_B \Omega
_{iC} + \frac{i}{4}
\bar{\Psi }_\mu {}^i \gamma ^\mu  \varepsilon _{ijk} \bar{z}^j{}_A
\bar{z}^k{}_B \lambda _C + \nonumber
\\
 && + \frac{i}{4\sqrt{2}} \bar{\Psi }_\mu {}^i \gamma ^\mu  z_i{}^A (z^B
\bar{z}_C) \rho  +
\frac{1}{2} \varepsilon ^{ijk} \bar{\Omega }_{iA} z_j{}^B \Omega _{kC} +
\bar{\Omega }_{iA}
\bar{z}^i{}_B \lambda _C + \nonumber \\
 && + \frac{1}{4} \bar{\Omega }_{iD} (z^A \bar{z}_B) \bar{z}^i{}_C \lambda
^D +
\frac{1}{2\sqrt{2}} \varepsilon ^{ijk} \bar{\Omega }_{iA} z_j{}^B z_k{}^C
\rho  + \nonumber
\\
 && + \frac{1}{2\sqrt{2}} \bar{\lambda }^A (z^B \bar{z}_C) \rho  \} +
\frac{1}{4}
(f^{ABC} z_i{}^A \bar{z}^j{}_B) (f^{CDF} z_j{}^D \bar{z}^i{}_F ) -
\nonumber \\
 && - \frac{1}{8} (f^{ABC} z^B \bar{z}_C)^2 + \frac{1}{8} (f z \bar{z}
\bar{z}^i) (f z_i z \bar{z}) \label{n37}
\end{eqnarray}
and to the fermion supertransformations, respectively,
\begin{eqnarray}
 \delta '\Psi _{\mu i} &=& - \frac{i}{8} \gamma _\mu  \left[ \varepsilon
^{ijk}(f z_j z_k \bar{z}^l)
+ (i\leftrightarrow l) \right] \eta _l \nonumber \\
 \delta '\Omega _{iA} &=& \left[ f^{ABC} z_i{}^B \bar{z}^j{}_C +
\frac{1}{2}
\bar{z}^k{}_A (f z_k z_i \bar{z}^j) \right] \eta _j - \nonumber   \\
 && - \frac{1}{2} \left[ f^{ABC} z^B \bar{z}_C + \frac{1}{2}
\bar{z}^k{}_A (f z_k z \bar{z}) \right] \eta _i \label{n38} \\
 \delta '\lambda ^A &=& \frac{1}{2} \varepsilon ^{ijk} \left[ f^{ABC}
z_i{}^B z_j{}^C +
\frac{1}{2} z_l{}^A (f z_i z_j \bar{z}^l) \right] \eta _k  \nonumber \\
 \delta '\rho  &=& - \frac{1}{2\sqrt{2}} (f z \bar{z} \bar{z}^i) \eta _i
\nonumber
\end{eqnarray}
where, for example, $(f z_i z_j \bar{z}^k) = f^{ABC} z_i{}^{A} z_j{}^B
\bar{z}^k{}_C$.

   As the global symmetry group $O(3,m)$ is non-compact, there exist a
lot of possible choices for the gauge group. The first class consists
of the groups like $O(3) \otimes  H$, $O(3,1) \otimes  H$, $O(2,1) \otimes
 O(1,2) \otimes  H$,
where $H$ is a compact group. Note, that in the case of $O(3,1)$ group
two variants are possible depending on what generators (compact or
non-compact) vector fields of supergravity multiplets correspond to
(that also determines the sign of cosmological term). In all the cases
considered the spontaneous supersymmetry breaking was accompanied by
the appearance of a cosmological term. Only in the last case imposing
a relation between two gauge coupling constants one can fine-tune
the value of cosmological constant to zero, but the scalar field
potential turns out to be non-stable.

   The second possibility is the nontrivial gauging similar to the
ones considered in \cite{Por89} for the case of $N=4$ supergravity.
The hidden sector for such model includes only one matter vector
multiplets and the gauge group is $E_2 \subset  O(3,1)$ (rotation and two
translations). The generators can be easily constructed from the
$O(3,1)$ ones in such a way that the appropriate coupling constants
are fully antisymmetric:
\begin{eqnarray}
\tau ^1 &=& \left( \begin{array}{ccc|c} 0 & 0 & 0 & 0 \\ 0 & 0 & 1 & 0 \\
0 & -1 & 0 & 0 \\ \hline 0 & 0 & 0 & 0 \end{array} \right) \qquad
\tau ^2 = \left( \begin{array}{ccc|c} 0 & 0 & -1 & 0 \\ 0 & 0 & 0 & 0 \\
1 & 0 & 0 & -1 \\ \hline 0 & 0 & -1 & 0 \end{array} \right) \nonumber
\\
\tau ^3 &=& \left( \begin{array}{ccc|c} 0 & 1 & 0 & 0 \\ -1 & 0 & 0 & 1 \\
0 & 0 & 0 & 0 \\ \hline 0 & 1 & 0 & 0 \end{array} \right) \qquad
\tau ^4 = \left( \begin{array}{ccc|c} 0 & 0 & 0 & 0 \\ 0 & 0 & -1 & 0 \\
0 & 1 & 0 & 0 \\ \hline 0 & 0 & 0 & 0 \end{array} \right)
\end{eqnarray}
The investigation of the scalar field potential shows that it has no
stable minima (the value of the cosmological term is indeed zero, but
it corresponds to the infinite vacuum expectation value of the
appropriate Higgs field).

   The last possibility is the pure "abelian" gauging corresponding to
three translations which was constructed in \cite{Zin91}. The hidden
sector for this model includes three matter vector multiplets and the
generators (which are the combinations of compact and non-compact ones
of the group $O(3,3)$) have the form:
\begin{equation}
 t^a = \left( \begin{array}{cc} \tau ^a & \tau ^a \\ -\tau ^a & -\tau ^a
\end{array}
\right)
\end{equation}
where $\tau ^a$, $a=1,2,3$ are usual antisymmetric $O(3)$ matrices. One
can easily check that the generators $t^a$ are nilpotent, i.e. $t^a
t^b = 0 \quad \forall  a,b$. This model really gives the only (as far as
we
know) example of the spontaneous supersymmetry breaking without a
cosmological term for the $N=3$ supergravity with matter. Due to the
requirement that structure constants to be antisymmetric one has only
one gauge coupling constant for all three translations and as a
consequence all three gravitini have equal masses.

   Thus we have seen that it seems impossible to have "realistic"
models in this class of theories. By term realistic we mean not only
the absence of a cosmological term, but also the possibility to have
spontaneous supersymmetry breaking with at least two different scales,
including partial super-Higgs effect $N=3 \to  N=1$. This result depends
heavily on the properties of the non-linear $\sigma $-model as well as on
the
global symmetry group $O(3,m)$. As we have already mentioned the model
$SU(3,m)/SU(3) \otimes  SU(m) \otimes  U(1)$ appears to be the only
candidate for
the $N=3$ supergravity-matter system, but the global symmetry groups
may drastically differ in dual versions. In \cite{Zin86b} there was
constructed a hidden sector for $N=3$ supergravity with global
symmetry group $GL(3,C) \otimes  T_9$ which turns out to be the dual
versions
for the $N=3$ supergravity with three vector multiplets. It was shown
that in such hidden sector spontaneous supersymmetry breaking with
three different scales are possible, in this the cosmological term is
equal to zero for all values of these parameters. So it would be
interesting to have a generalization of this hidden sector to the case
of arbitrary number of vector multiplets. In our next section we are
going to construct such a model.

\section{Dual version}

   The crucial observation for the whole construction is that since
the dual versions have the same scalar field geometry one can use all
the terms in the Lagrangian and in the supertransformations without
vector fields obtained previously. To construct the desired
generalization of the hidden sector \cite{Zin86b} we start by solving
scalar field constraint (\ref{c1}). For that purpose we introduce a
kind of "light cone" variables $z_i{}^A = (x_i{}^\alpha  + y_i{}^\alpha ,
x_i{}^\alpha
- y_i{}^\alpha , z_i{}^a)$, $i,\alpha  = 1,2,3$, $a=7,8...m+3$. Then the
constraint (\ref{c1}) takes the form:
\begin{equation}
 x_i{}^\alpha  \bar{y}_\alpha {}^j + y_i{}^\alpha  \bar{x}_\alpha {}^j =
\delta _i^j + \frac{1}{2}
z_i{}^a \bar{z}_a{}^j.
\end{equation}
Now we introduce a new variable $\pi _{\alpha }{}^{\beta
}=(y^{-1})_{\alpha }{}^i
x_i{}^{\beta }-\bar{x}_{\alpha }{}^i(\bar{y}^{-1})_i{}^{\beta }$ (such a
special
choice will become clear later) so that we can exclude the field
$x_i{}^\alpha $ and rewrite all the formulas in terms of $y_i{}^{\alpha
}$,
$\pi _{\alpha }{}^{\beta }$ and $z_i{}^a$. In this, it appears that in
order to have
a canonical form for the scalar field kinetic terms, one have to make
a change $z_i{}^a = y_i{}^\alpha  z_\alpha {}^a$. As a result we obtain:
\begin{eqnarray}
 && \frac{1}{2} D_\mu  z_i{}^a D^\mu  \bar{z}_a{}^i = \frac{1}{4} Sp
\left[
(\partial _\mu  y (y^{-1}) + (\bar{y}^{-1}) \partial _\mu  \bar{y})^2+
\right. \nonumber \\
 && \left. + 2 \bar{y} y \partial _\mu  z \partial _\mu  \bar{z} - [y
(\partial _\mu  \pi  + \frac{1}{2}
(z \stackrel{\leftrightarrow }{\partial }_\mu  \bar{z})) \bar{y}]^2
\right].
\end{eqnarray}

   Analogously, to solve the spinor field constraint (\ref{c2}) we
introduce new variables:
\begin{equation}
 \Omega _{iA} = (-(\chi _{i\alpha } + \eta _{i\alpha }), (\eta _{i\alpha }
- \chi _{i\alpha }), \Omega _{ia}), \quad
\lambda ^A = ((\xi ^\alpha  + \lambda ^\alpha ), (\xi ^\alpha  - \lambda
^\alpha ), \Lambda ^a).
\end{equation}
The following changes of variables are necessary to bring the spinor
fields kinetic terms to the canonical form:
\begin{eqnarray}
 && \chi _{i\alpha } \to  \frac{1}{\sqrt{2}} \bar{y}_\alpha {}^i \chi
_{ij}, \qquad \Omega _{ia}
\to  \Omega _{ia} + \frac{1}{\sqrt{2}} \bar{z}_a{}^\alpha  \bar{y}_\alpha
{}^j \chi _{ij}
\nonumber \\
 && \lambda ^\alpha  \to  \frac{1}{\sqrt{2}} y_i{}^\alpha  \lambda ^i,
\qquad  \Lambda ^a \to  \Lambda ^a +
\frac{1}{\sqrt{2}} z_\alpha ^a y_i{}^\alpha  \lambda ^i.
\end{eqnarray}

   Let us discard for a moment "matter" fields $z_\alpha {}^a$, $\Omega
_{ia}$ and
$\Lambda ^a$. Then all the terms in the Lagrangian without vector fields
in
terms of new variables take the form:
\begin{eqnarray}
 L_1 &=& - \frac{1}{2} R + \frac{i}{2} \varepsilon ^{\mu \nu \rho \sigma }
\bar{\Psi }_{\mu i} \gamma _5
\gamma _\nu  D_\rho  \Psi _{\sigma i} + \frac{i}{2} \bar{\chi }_{ij}
\hat{D} \chi _{ij} +
\frac{i}{2} \bar{\rho } \hat{D} \rho  + \nonumber  \\
 && + \frac{1}{4} Sp\{(S_\mu )^{2} - (U_\mu )^{2}\} + \frac{1}{2\sqrt{2}}
 \varepsilon ^{ijk} \bar{\chi }_{il} \gamma ^\mu  \gamma ^\nu  (S_\nu  -
U_\nu )_j{}^l \Psi _{\mu k} - \nonumber \\
 && - \frac{1}{2\sqrt{2}} \bar{\lambda }^i \gamma _\mu  \gamma _\nu
(S_\nu  + U_\nu )_i{}^j \Psi _{\mu j}
- \frac{i}{4} \bar{\chi }_{ij} \gamma ^\mu  (P_\mu  + U_\mu )_i{}^k \chi
_{kj} - \nonumber \\
 && - \frac{i}{4} \bar{\chi }_{ij} \gamma ^\mu  (P_\mu  - U_\mu )_j{}^k
\chi _{ik} +
\frac{i}{8} \bar{\chi }_{ij} \gamma ^\mu  U_\mu  \chi _{ij} + \frac{i}{4}
\bar{\lambda }^i \gamma ^\mu
(P_\mu  - U_\mu )_j{}^i \lambda ^j - \nonumber \\
 && - \frac{i}{8} \bar{\lambda }^i \gamma ^\mu  U_\mu  \lambda ^i -
\frac{i}{4} \varepsilon ^{\mu \nu \rho \sigma }
\bar{\Psi }_{\mu i} \gamma _5 \gamma _\nu  \{(U_\rho  + P_\rho )_i{}^j -
\frac{1}{2} U_\rho  \delta _i{}^j
\}\Psi _{\sigma j} \label{l1}
\end{eqnarray}
where we introduced the following notations:
\begin{eqnarray}
 S_{\mu i}{}^j &=& \{\partial _\mu  y (y^{-1}) + (\bar{y}^{-1}) \partial
_\mu  \bar{y}
\}_i{}^j,  \nonumber \\
 P_{\mu i}{}^j &=& \{\partial _\mu  y (y^{-1}) - (\bar{y}^{-1}) \partial
_\mu  \bar{y}
\}_i{}^j,   \\
 U_{\mu i}{}^j &=& \{y \partial _\mu  \pi  \bar{y}\}_i{}^j, \qquad U_\mu
= U_{\mu i}{}^i
 \nonumber
\end{eqnarray}
while the appropriate terms in the supertransformations look like:
\begin{eqnarray}
 \delta e_{\mu r} &=& i(\bar{\Psi }_\mu {}^i \gamma _r \eta _i) \nonumber
\\
 \delta \Psi _{\mu i} &=& 2D_\mu  \eta _i - \{(P_\mu  + U_\mu )_i{}^j -
\frac{1}{2} \delta _i{}^j
U_\mu \} \eta _j \nonumber \\
 \delta \chi _{ij} &=& \frac{i}{\sqrt{2}} \varepsilon _{ikl} \gamma ^\mu
(S_\mu  - U_\mu )_k{}^j \eta _l
\qquad  \delta \lambda ^i = - \frac{i}{\sqrt{2}} \gamma ^\mu  (S_\mu  +
U_\mu )_i{}^j \eta _j
\label{s1} \\
 \delta \pi _{\alpha }{}^{\beta } &=& - \sqrt{2} (\bar{\lambda }^i
(\bar{y}^{-1})_i{}^{\beta }
(y^{-1})_{\alpha }{}^j \eta _j) + \sqrt{2} \varepsilon ^{ijk} (\bar{\chi
}_{jl}
(\bar{y}^{-1})_i{}^{\beta } (y^{-1})_{\alpha }{}^l \eta _k) \nonumber \\
 \delta \varphi _i{}^\alpha  &=& \frac{1}{\sqrt{2}} \varepsilon ^{ijk}
(\bar{\chi }_{jl} \bar{y}_\alpha {}^l
\eta _k) + \frac{1}{\sqrt{2}} (\bar{\lambda }^j y_j{}^\alpha  \eta _i)
\nonumber \\
 \delta \pi _i{}^\alpha  &=& \frac{1}{\sqrt{2}} \varepsilon ^{ijk}
(\bar{\chi }_{jl} \gamma _5
\bar{y}_\alpha {}^l \eta _k) - \frac{1}{\sqrt{2}} (\bar{\lambda }^j \gamma
_5 y_j{}^\alpha  \eta _i)
\nonumber
\end{eqnarray}
One can see that (up to slightly different notations) this formulas
indeed coincide with the corresponding ones from the \cite{Zin86b}. In
this, the complex fields $y_i{}^\alpha $ describes a non-linear $\sigma
$-model
$GL(3,C)/U(3)$, while the fields $\pi _\alpha {}^\beta $ enter the
Lagrangian
through the divergency $\partial _\mu  \pi _\alpha {}^\beta $ only (thus
explaining our special
choice for it) so that the Lagrangian is invariant under the global
translations $\pi _\alpha {}^\beta  \to  \pi _\alpha {}^\beta  + \Lambda
_\alpha {}^\beta $.

   In our current notations the part of the hidden sector containing
vector fields $V_{\mu \alpha } = A_{\mu \alpha } + \gamma _5 B_{\mu \alpha
}$, $\alpha =1,2,3$ has the form:
\begin{eqnarray}
L_2 &=& - \frac{1}{4} (y^{-1})_\alpha {}^i (\bar{y}^{-1})_i{}^\beta
(V^*_{\mu \nu })^\alpha  (V^{\mu \nu })_\beta  - \frac{i}{4} \pi _\alpha
^\beta  (\tilde{V}^*_{\mu \nu })^\alpha
(V_{\mu \nu })_\beta  + \frac{i}{2} \bar{\rho } \hat{D} \rho  + \nonumber
\\
 && + \frac{i}{8} \bar{\rho } \gamma ^\mu  U_\mu  \rho  + \frac{1}{4}
\varepsilon ^{ijk} \bar{\Psi }_{\mu i}
(\bar{y}^{-1})_j{}^\alpha  (V^{\mu \nu } - \gamma _5 \tilde{V}^{\mu \nu
})_\alpha  \Psi _{\nu k} +
\nonumber \\
 && + \frac{i}{4\sqrt{2}} \bar{\rho } \gamma ^\mu  (\sigma V^*)^\alpha
(y^{-1})_\alpha {}^i \Psi _{\mu i} +
 \frac{i}{4\sqrt{2}} \bar{\chi }_{ij} \gamma ^\mu  (\sigma V)_\alpha
(\bar{y}^{-1})_j{}^\alpha
\Psi _{\mu i} \nonumber \\
 && - \frac{1}{4} \bar{\lambda }^j (\sigma V^*)^\alpha  (y^{-1})_\alpha
{}^i \chi _{ij} - \bar{\lambda }^i
(\bar{y}^{-1})_i{}^\alpha  (\sigma V)_\alpha  \rho  \label{l2}
\end{eqnarray}
and the corresponding terms in the supertransformations:
\begin{eqnarray}
 \delta \Psi _{\mu i} &=& \frac{i}{4} \varepsilon _{ijk} (y^{-1})_\alpha
{}^j (\sigma V^*)^\alpha  \gamma _\mu  \eta ^k
\nonumber \\
 \delta A_\mu {}^\alpha  &=& - \varepsilon ^{ijk}(\bar{\Psi }_{\mu i}
y_j{}^\alpha  \eta _k) + \frac{1}{\sqrt{2}}
(\bar{\chi }_{ij} \gamma _\mu  y_j{}^\alpha  \eta _i) + \frac{i}{\sqrt{2}}
(\bar{\rho } \gamma _\mu
\bar{y}_\alpha {}^i \eta _i) \nonumber  \\
 \delta B_\mu {}^\alpha  &=& - \varepsilon ^{ijk}(\bar{\Psi }_{\mu i}
\gamma _5 y_j{}^\alpha  \eta _k) +
\frac{1}{\sqrt{2}} (\bar{\chi }_{ij} \gamma _\mu  \gamma _5 y_j{}^\alpha
\eta _i) -
\frac{i}{\sqrt{2}} (\bar{\rho } \gamma _\mu  \gamma _5 \bar{y}_\alpha {}^i
\eta _i) \label{s2}
\\
 \delta \chi _{ij} &=& - \frac{1}{2\sqrt{2}} (\sigma V)_\alpha
(\bar{y}^{-1})_j{}^\alpha  \eta _i
\qquad  \delta \rho  = - \frac{1}{2\sqrt{2}} (\sigma V^*)^\alpha
(y^{-1})_\alpha {}^j \eta _i
\nonumber
\end{eqnarray}

   Now we can restore all the terms with the matter fields $z_\alpha
{}^a$,
$\Omega _{i\alpha }$ and $\Lambda ^a$ in the Lagrangian:
\begin{eqnarray}
 L_3 &=& \frac{1}{2} y_i{}^\alpha  \partial _\mu  z_\alpha {}^a \partial
_\mu  \bar{z}_\beta {}^a
\bar{y}_\beta {}^i + \frac{i}{2} \bar{\Omega }_{ia} \hat{D} \Omega _{ia} +
\frac{i}{2}
\bar{\Lambda }^a \hat{D} \Lambda ^a - \nonumber  \\
 && - \frac{i}{\sqrt{2}} \bar{\chi }_{ij} \bar{y}_\alpha {}^j
\hat{\partial } z_\alpha ^a
\Omega _{ia} - \frac{i}{4} \bar{\Omega }_{ia} \gamma ^\mu  (P_\mu  + U_\mu
)_i{}^j \Omega _{ja} +
\frac{i}{8} \bar{\Omega } \gamma ^\mu  U_\mu  \Omega _{ia} + \nonumber \\
 && + \frac{1}{2} \varepsilon ^{ijk} \bar{\Omega }_{ia} \gamma ^\mu
\gamma ^\nu  y_j{}^\alpha  \partial _\nu  z_\alpha ^a
\Psi _{\mu k} - \frac{i}{\sqrt{2}} \bar{\lambda }^i y_i{}^\alpha
\hat{\partial } \bar{z}_a^\alpha  \Lambda ^a
- \frac{i}{8} \bar{\Lambda }^a \gamma ^\mu  U_\mu  \Lambda ^a - \nonumber
\\
 && - \frac{1}{2} \bar{\Lambda }^a \gamma ^\mu  \gamma ^\nu
\bar{y}_\alpha {}^i \partial _\nu  \bar{z}_a^\alpha
\Psi _{\mu i}  \label{l3}
\end{eqnarray}
as well as in the supertransformations
\begin{eqnarray}
 \delta \Lambda ^a &=& -i \hat{\partial } \bar{z}_a^\alpha  \bar{y}_\alpha
{}^i \eta _i \qquad
 \delta \Omega _{ia} = i\varepsilon _{ijk} \hat{\partial } z_\alpha ^a
y_j{}^\alpha  \eta _k  \nonumber \\
 \delta \pi _\alpha {}^\beta  &=& (\bar{\Lambda }^a \bar{z}_a^\beta
(y^{-1})_\alpha {}^i \eta _i) - \varepsilon ^{ijk}
(\bar{\Omega }_{ja} z_\alpha ^a (\bar{y}^{-1})_i{}^\beta  \eta _k)
\label{s3} \\
 \delta \varphi _a{}^\alpha  &=& \varepsilon ^{ijk} (\bar{\Omega }_{ja}
(\bar{y}^{-1})_i^\alpha  \eta _k) +
(\bar{\Lambda }^a (y^{-1})_\alpha {}^i \eta _i) \nonumber \\
 \delta \pi _a{}^\alpha  &=& \varepsilon ^{ijk} (\bar{\Omega }_{ja} \gamma
_5 (\bar{y}^{-1})_i^\alpha  \eta _k) -
(\bar{\Lambda }^a \gamma _5 (y^{-1})_\alpha {}^i \eta _i) \nonumber
\end{eqnarray}
Note that in the presence of matter fields the expression for the
$U_{\mu i}{}^j$ has the form  $U_{\mu i}{}^j = \{ y [\partial _\mu  \pi  +
\frac{1}{2}
z \stackrel{\leftrightarrow }{\partial }_\mu  \bar{z}] \bar{y} \}$.

   Let us stress ones more that all the terms without vector fields
are exactly the same in all dual versions. Now it is a relatively easy
task to complete the whole construction to include matter vector
fields $C_\mu {}^a$ by the use of the usual Noether procedure. That give
the following additional terms the Lagrangian:
\begin{eqnarray}
 L_4 &=& - \frac{1}{8} z_\alpha ^a (V^*_{\mu \nu })^\alpha
\bar{z}_a^\beta  (V_{\mu \nu })_\beta  -
\frac{1}{16} \left\{z_\alpha ^a (V^*_{\mu \nu })^\alpha  z_\beta ^a
(V^*_{\mu \nu } +
i\tilde{V}^*_{\mu \nu })^\beta  + h.c. \right\} + \nonumber \\
 && + \frac{1}{4} \left\{ z_\alpha ^a C^a_{\mu \nu } (V^*_{\mu \nu } +
i\tilde{V}^*_{\mu \nu })^\alpha  + h.c. \right\} - \frac{1}{4} (C^a_{\mu
\nu })^2 +
\nonumber \\
 && + \frac{1}{4} \bar{\Omega }_{ia} (\sigma V^*)^\alpha  (y^{-1})_\alpha
{}_i \Lambda ^a -
\frac{1}{2\sqrt{2}} \bar{\Lambda }^a (\sigma C)^a \rho  + \frac{i}{4}
\bar{\Omega }_{ia}
\gamma ^\mu  (\sigma C)^a \Psi _{\mu i} - \label{l4} \\
 && - \frac{i}{8} \bar{\Omega }_{ia} \gamma ^\mu  [\bar{z}_a^\alpha
(\sigma V)_\alpha  + z_\alpha ^a
(\sigma V^*)^\alpha ] \Psi _{\mu i} + \frac{1}{4\sqrt{2}} \bar{\Lambda }^a
[z_\alpha ^a (\sigma V^*)^\alpha  +
\bar{z}_a^\alpha  (\sigma V)_\alpha ] \rho   \nonumber
\end{eqnarray}
and to the supertransformations:
\begin{eqnarray}
 \delta \Omega _{ia} &=& \frac{1}{4} \{\bar{z}_a^\alpha  (\sigma V)_\alpha
 + z_\alpha ^a (\sigma V^*)^\alpha  \} \eta _i
- \frac{1}{2} (\sigma C)^a \eta _i \nonumber \\
 \delta C^a_\mu  &=& i(\bar{\Omega }_{ia} \gamma _\mu  \eta _i) +
\frac{i}{\sqrt{2}} (\bar{\chi }_{ij}
\gamma _\mu  y_j{}^\alpha  z_\alpha ^a \eta _i) + \label{s4} \\
 && + \frac{i}{\sqrt{2}} (\bar{\rho } \gamma _\mu  \bar{y}_\alpha ^i
\bar{z}_a^\alpha  \eta _i) -
\varepsilon ^{ijk} (\bar{\Psi }_{\mu i} y_j{}^\alpha  z_\alpha ^a \eta _k)
  \nonumber
\end{eqnarray}

\section{Spontaneous symmetry breaking}

   As we have already noted, in extended supergravities the appearance
of the mass terms is tightly connected with the switching of gauge
interaction. We have seen that all the attempts to get spontaneous
supersymmetry breaking through the non-abelian gauge interaction
usually lead to the appearance of the cosmological term. In the dual
version constructed in the previous section we have fields $\pi _\alpha
{}^\beta $
that enter the Lagrangian through the divergency $\partial _\mu  \pi
_\alpha {}^\beta $ only,
so that the Lagrangian is invariant under the  global translations
 $\pi _\alpha {}^\beta  \to  \pi _\alpha {}^\beta  + \Lambda _\alpha
{}^\beta $. In \cite{Zin86b} it was shown that
using these fields as a Goldstone ones we may introduce the vector
field masses by going to the local transformations and making a
change  $\partial _\mu  \pi _{\alpha }{}^{\beta } \to  \partial _\mu  \pi
_{\alpha }{}^{\beta } - (V_\mu )_\alpha {}^\beta $, where
\begin{equation}
 (V_\mu )_\alpha {}^\beta  = (\Sigma ^\gamma )_\alpha {}^\beta  (V_\mu
)_\gamma  + (\Sigma _\gamma )_\alpha {}^\beta  (V^{*}_\mu )^\gamma
\end{equation}
Here we have introduce matrices
\begin{eqnarray}
 (\Sigma ^1) &=& \left( \begin{array}{ccc} 0 & 0 & 0 \\ 0 & 0 & \mu _2
 \\ 0 & \mu _3 & 0 \end{array} \right) \quad (\Sigma ^2) = \left(
\begin{array}{ccc}  0 & 0 & \mu _1 \\ 0 & 0 & 0 \\ -\mu _3 & 0 & 0
\end{array}
\right) \quad (\Sigma ^3) = \left( \begin{array}{ccc} 0 & -\mu _1 & 0
\\ -\mu _2 & 0 & 0 \\ 0 & 0 & 0 \end{array} \right) \nonumber \\
 (\Sigma _\gamma ) &=& - (\Sigma ^\gamma )^T ,\quad (\Sigma _\gamma
)_\alpha {}^\beta  = -(\Sigma _\alpha )_\gamma {}^\beta , \quad
(\Sigma ^\gamma )_\alpha {}^\beta  = - (\Sigma ^\beta )_\alpha {}^\gamma
\label{p1}
\end{eqnarray}
All the terms in the Lagrangians (\ref{l1}), (\ref{l2}) and in the
supertransformations (\ref{s1}), (\ref{s2}) containing $\partial _\mu  \pi
_\alpha {}^\beta $
are covariantized just by the change  $\partial _\mu  \pi _\alpha {}^\beta
 \to  \partial _\mu
\pi _\alpha {}^\beta  - (V_\mu )_\alpha {}^\beta $ except the "axionic"
term $-\frac{i}{4}
\pi _\alpha {}^\beta  (V^{\mu \nu })_\beta  (\tilde{V}_{\mu \nu
}^*)^\alpha $ which have to be completed
with $-\frac{1}{6} (\tilde{V}^{\mu \nu })^\alpha  (V_\mu )_\beta  (V_\mu
)_\alpha {}^\beta  + h.c.$.
Thus the requirement of gauge invariance leads to the self-interaction
for the abelian vector fields!

   As usual, the transition from the global to local transformations
and the change of the usual derivatives by gauge covariant ones spoils
the invariance of the Lagrangian under the supertransformatrions. In
order to restore such invariance it is  necessary to add to the
Lagrangian the following terms:
\begin{eqnarray}
 L' &=& \frac{1}{4} (\bar{\Psi }_{\mu i} \sigma ^{\mu \nu } \varepsilon
^{ijk} (\Sigma _j)_k{}^l \Psi _{\nu l}) +
\frac{i}{\sqrt{2}} \bar{\Psi }_{\mu i} \gamma ^\mu  (\Sigma ^j)_i{}^k \chi
_{jk}) -
\frac{i}{2\sqrt{2}} (\bar{\Psi }_{\mu i} \gamma ^\mu  \Sigma ^j \chi
_{ij}) - \nonumber \\
 && - \frac{i}{2\sqrt{2}} (\bar{\Psi }_{\mu i} \gamma ^\mu  \Sigma _i \rho
) -
\frac{i}{2\sqrt{2}}  (\bar{\Psi }_{\mu i} \gamma ^\mu  \varepsilon _{ijk}
(\Sigma ^j)_l{}^k \lambda ^l) +
\nonumber \\
 && - (\frac{1}{2} \bar{\chi }_{ij} \varepsilon ^{ikm} (\Sigma ^j)_m{}^l
\chi _{kl}) +
(\bar{\lambda }^i (\Sigma ^j)_i{}^k \chi _{jk}) + \frac{1}{2}
(\bar{\lambda }^i \Sigma ^j \chi _{ji}) +
\nonumber \\
 && + \frac{1}{2} (\bar{\lambda }^i \Sigma _i \rho ) + \frac{1}{2}
(\bar{\chi }_{ij}
\varepsilon ^{kli} (\Sigma _k)_l{}^j \rho ) + \frac{1}{2} (\Sigma _i)
(\Sigma ^i) \label{l5}
\end{eqnarray}
where
\begin{eqnarray}
 (\Sigma _i)_j{}^k &=& y_i{}^\alpha  y_j{}^\beta  (\Sigma _\alpha )_\beta
{}^\gamma  \bar{y}_\gamma {}^k \qquad
\Sigma _i = (\Sigma _i)_j{}^j  \label{p2}
\end{eqnarray}
In this, the whole Lagrangian (\ref{l1})+(\ref{l2})+(\ref{l5}),
describing hidden sector of our dual version will be invariant under
the supertransformations (\ref{s1})+(\ref{s2}) with the additional
terms:
\begin{eqnarray}
 \delta '\Psi _{\mu i} &=& - \frac{i}{4} \gamma _\mu  \left[ \varepsilon
^{ijk} (\Sigma _j)_k{}^l + (i \leftrightarrow l)
\right] \eta _l \qquad \delta '\rho  = - \frac{1}{\sqrt{2}} \Sigma ^i \eta
_i  \nonumber \\
 \delta '\chi _{ij} &=& - \sqrt{2} (\Sigma _i)_j{}^k \eta _k -
\frac{1}{\sqrt{2}} \Sigma _j \eta _i
\qquad \delta '\lambda ^i = \frac{1}{\sqrt{2}} \varepsilon ^{ljk} (\Sigma
_j)_k{}^i \eta _l  \label{s5}
\end{eqnarray}

    From the formula (\ref{l5}) one can see that the scalar field
potential in such model has the form $V = - 1/2 \Sigma _i \Sigma ^i$.
Using
(\ref{p1}) and (\ref{p2}) it can be shown that this potential is
positively defined and its value at the minimum is equals to zero,
which corresponds to the absence of the cosmological term. Note, that
at the minimum vacuum expectation value for $y_i{}^\alpha $ is $\delta
_i{}^\alpha $.
At the same time, the gravitini mass matrix has the form
\begin{eqnarray}
 \frac{1}{2} \varepsilon ^{\alpha \gamma \delta } (\Sigma _\gamma )_\delta
{}^\beta  &=& \left(\begin{array}{ccc} -\mu _1 &
0 & 0 \\ 0 & \mu _2 & 0 \\ 0 & 0 & -\mu _3 \end{array} \right)^{\alpha
\beta }
\end{eqnarray}
Thus, in such model there exist a possibility to have spontaneous
supersymmetry breaking with three arbitrary scales, including partial
super-Higgs effect  $N = 3 \to  N =2$ ¨ $N = 3 \to  N = 1$. Let us stress
ones more that the cosmological term is absent for any values of these
parameters.

    When the matter fields are present there exist a possibility to
add gauge interaction for matter vector multiplets with any compact
gauge group $H$. Such a gauging gives additional (positively defined)
contribution to the scalar field potential but doesn't change the main
properties of our model. For simplicity we will not consider these
additional terms here. In this case, the only new term that we have to
add to the Lagrangian is:
\begin{equation}
 L''= \frac{1}{2} (\bar{\Lambda }^a \Sigma ^i \Omega _{ia})
\end{equation}
This term does not lead to the mass generation for matter spinors $\Omega
$
and $\Lambda $ because the vacuum expectation value for the $\Sigma ^i$ is
equal to
zero.

\section{Conclusion}

   Thus, we have seen that in the "minimal" model for $N=3$
supergravity with vector multiplets the only possibility for
spontaneous supersymmetry breaking without a cosmological term is the
model \cite{Zin91} in which one have only one breaking scale and all
three gravitini are mass degenerate. At the same time in the dual
version the spontaneous supersymmetry breaking with three different
scales (including partial super-Higgs effect $N=3 \to  N=1$) is possible,
but no soft breaking terms for the matter fields are generated. In
our subsequent paper we have shown that the situation in $N=4$
supergravity is to much extent similar, but with some more interesting
possibilities.

\hskip 1cm

{\bf Acknowledgments}

\hskip 0.5cm

Work supported by International Science Foundation grant RMP000 and by
Russian Foundation for Fundamental Research grant 94-02-03552.

\end{document}